\documentclass[%
 reprint,
 superscriptaddress,
 amsmath,amssymb,
aps,
prb,
citeautoscript
]{revtex4-1}
\usepackage[normalem]{ulem}
\usepackage{natbib}
\usepackage{xcolor}
\usepackage{graphicx}
\usepackage{amsmath}
\usepackage{physics}
\usepackage[colorlinks=true,allcolors=blue]{hyperref}%
\DeclareUnicodeCharacter{2212}{-}

\usepackage{xspace}

\renewcommand{\sout}[1]{\unskip}
\newcommand{\figref}[2]{\hyperref[#1]{Figure~\ref{#1}#2}}
\newcommand{\figureref}[2]{\hyperref[#1]{Figure~\ref{#1}#2}}

\usepackage{nameref}
\usepackage{hyperref}
\hypersetup{colorlinks,allcolors=blue}

\begin{document}

\title{Chiral Altermagnon in MnTe}

\date{\today}
\author{Daniel~Jost}
\email{daniel.jost@stanford.edu}
\affiliation{Stanford Institute for Material and Energy Science (SIMES), SLAC National Accelerator Laboratory, Menlo Park, CA, USA}
\affiliation{Linac Coherent Light Source, SLAC National Accelerator Laboratory, Menlo Park, CA, USA}

\author{Resham~B.~Regmi}
\affiliation{Department of Physics and Astronomy, University of Notre Dame, Notre Dame, IN 46556, USA}
\affiliation{Stravropoulos Center for Complex Quantum Matter,
University of Notre Dame, Notre Dame, IN 46556, USA}

\author{Eder~G.~Lomeli}
\affiliation{Department of Materials Science and Engineering, Stanford University, Stanford, CA, USA}
\affiliation{Stanford Institute for Material and Energy Science (SIMES), SLAC National Accelerator Laboratory, Menlo Park, CA, USA}

\author{Samuel~Sahel-Schackis}
\affiliation{Stanford PULSE Institute, SLAC National Accelerator Laboratory, Menlo Park, CA, USA}
\affiliation{Linac Coherent Light Source, SLAC National Accelerator Laboratory, Menlo Park, CA, USA}
\affiliation{Department of Physics, Stanford University, Stanford, CA, 94305, USA}

\author{Monika~Scheufele}
\affiliation{Walther-Meißner-Institut, Bayerische Akademie der Wissenschaften, 85748 Garching,
Germany}
\affiliation{Technical University of Munich, TUM School of Natural Sciences, Physics Department, Garching, Germany}

\author{Marcel~Neuhaus}
\affiliation{Linac Coherent Light Source, SLAC National Accelerator Laboratory, Menlo Park, CA, USA}

\author{Rachel~Nickel}
\affiliation{European Synchrotron Radiation Facility, BP 220, F-38043, Grenoble Cedex, France
}

\author{Flora~Yakhou}
\affiliation{European Synchrotron Radiation Facility, BP 220, F-38043, Grenoble Cedex, France
}

\author{Kurt~Kummer}
\affiliation{European Synchrotron Radiation Facility, BP 220, F-38043, Grenoble Cedex, France
}

\author{Nicholas~Brookes}
\affiliation{European Synchrotron Radiation Facility, BP 220, F-38043, Grenoble Cedex, France
}

\author{Lingjia~Shen}
\affiliation{Linac Coherent Light Source, SLAC National Accelerator Laboratory, Menlo Park, CA, USA}

\author{Georgi~L.~Dakovski}
\affiliation{Linac Coherent Light Source, SLAC National Accelerator Laboratory, Menlo Park, CA, USA}

\author{Nirmal~J.~Ghimire}
\affiliation{Department of Physics and Astronomy, University of Notre Dame, Notre Dame, IN 46556, USA}
\affiliation{Stravropoulos Center for Complex Quantum Matter,
University of Notre Dame, Notre Dame, IN 46556, USA}

\author{Stephan~Gepr\"ags}
\affiliation{Walther-Meißner-Institut, Bayerische Akademie der Wissenschaften, 85748 Garching,
Germany}

\author{Matthias~F.~Kling}
\affiliation{Stanford PULSE Institute, SLAC National Accelerator Laboratory, Menlo Park, CA, USA}
\affiliation{Linac Coherent Light Source, SLAC National Accelerator Laboratory, Menlo Park, CA, USA}
\affiliation{Department of Applied Physics, Stanford University. Stanford, California 94305, USA}

\date{\today}

\begin{abstract}
Altermagnetism has surfaced as a novel magnetic phase, bridging the properties of ferro- and anti-ferromagnetism. The momentum-dependent spin-splitting observed in these materials reflects their unique symmetry characteristics, which also establish the conditions for chiral magnons to emerge. Here we provide the first direct experimental evidence for a chiral magnon in the altermagnetic candidate MnTe revealed by circular-dichroism resonant inelastic X-ray scattering (CD-RIXS). This mode which we term chiral altermagnon exhibits a distinct momentum dependence of its spin polarization consistent with the proposed altermagnetic $g-$wave symmetry of MnTe. Our polarization-resolved results corroborate the existence of a new class of magnetic excitations, demonstrating how altermagnetic order shapes spin dynamics and paves the way for advances in spintronic and quantum technologies.
\end{abstract}

\maketitle

\noindent

\section{Introduction}
Chiral magnons are collective modes associated with the magnetic spins in materials that exhibit a definite handedness with respect to their spin polarization. These excitations represent an emerging frontier, offering novel opportunities to enable spintronic devices and spin-based information processing~\cite{Smejkal:2018, Tokura:2019, Dieterle:2019, Yu:2019, Wang:2020, Yang:2020, Yang:2021, Ma:2021}. Traditionally, chiral magnons have been associated with magnets exhibiting either net magnetization or complex spin textures~\cite{Fert:2017, Zhou:2024}. The spin dynamics in these materials are biased by the net magnetization, producing collective modes with an intrinsic handedness and direction-dependent propagation. These features are nominally absent once spin sub-lattices are energetically degenerate, as in anti-ferromagnets. However, recent theoretical advances suggest that chiral magnons~\cite{Nambu:2020, Cheong:2022} may also arise in so called altermagnets~\cite{Smejkal:2020, Smejkal:2022:1, Smejkal:2022:2} that, despite having zero net magnetization like anti-ferromagnets, break spin symmetry. 
 
\begin{figure*} 
	\centering
	\includegraphics[width=180mm]{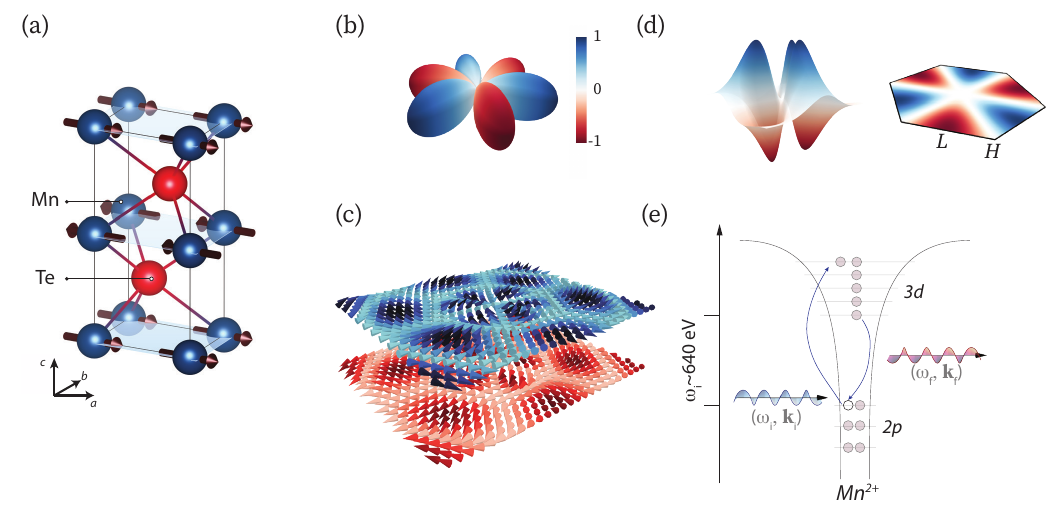} 
	\caption{\textbf{Crystalline structure, altermagnetic spin polarization as well as collective magnetic excitations probed by resonant inelastic X-ray scattering (RIXS).}  (a)~Primitive unit cell of MnTe showing the ferromagnetic spin ordering of the Mn-moments within each plane separated by non-magnetic Te atoms, and their anti-ferromagnetic stacking~\cite{Kunitomi:1964, Efrem:2005, Kriegner:2017}. Along with the magnetically inert Te spacer layers and the anti-ferromagnetic ordering of the Mn-moments,  the opposite-spin sublattices of MnTe are related by a rotation rather than by inversion or translation which is a key feature of altermagnets~\cite{Smejkal:2022:1,Smejkal:2022:2}. (b) Altermagnetic $g-$wave symmetry of the order parameter predicted for MnTe. The color gradient represents the spin polarization. (c) Collective magnetic excitations in real space represented as rotations of the spins sitting on each site of the hexagonal lattice. (d) Reciprocal space representation of the spin polarization. The right-hand image represents a projection onto the first Brillouin zone. (e) Schematic of the CD-RIXS process for a transition from Mn $2p$ core level to $3d$ orbitals which harbor an $S=5/2$ spin configuration and subsequent de-excitation, filling the core hole upon emitting a photon of energy $\omega = \omega_\mathrm{i}-\omega_\mathrm{f}$ and momentum $\mathbf{q} = \mathbf{k}_\mathrm{i}-\mathbf{k}_\mathrm{f}$.}
	\label{fig:Fig1} 
\end{figure*}
A proposed altermagnet, manganese telluride (MnTe)~\cite{Juza:1956} offers an ideal platform to explore this notion. While hosting robust $A-$type anti-ferromagnetic ordering [\figureref{fig:Fig1}{(a)}]~\cite{Kunitomi:1964, Efrem:2005, Kriegner:2017}, it displays electronic and magnetic properties~\cite{Krempasky:2024, Lee:2024, Osumi:2024, Hajlaoui:2024, Liu:2024, Amin:2024} unexpected for an anti-ferromagnet~\cite{Nagaosa:2010, Smejkal:2022_NatRevMat}. More fundamentally, the opposite-spin sublattices are related by a rotation rather than by inversion or translation which is a key feature of altermagnets~\cite{Smejkal:2022:1,Smejkal:2022:2}. As a consequence, crystalline anisotropies imprint on the spin structure and shape the symmetry of the altermagnetic order parameter in MnTe [\figureref{fig:Fig1}{(b),(c)}], characterized by a predicted altermagnetic $g-$wave symmetry of spin polarization in momentum space [\figureref{fig:Fig1}{(d)}]~\cite{Smejkal:2022:1}. This $g-$wave symmetry governs a momentum-dependent splitting of the electronic band structure, recently confirmed in photo-emission experiments~\cite{Krempasky:2024, Lee:2024, Osumi:2024, Hajlaoui:2024}. Similarly, collective magnetic excitations, also called magnons, which would nominally remain degenerate in anti-ferromagnets are expected to split in certain momentum-space regions~\cite{Smejkal:2023}. Initial observations suggest that such a splitting indeed occurs~\cite{Liu:2024}. Taken together, these results provide convincing evidence that MnTe realizes altermagnetism and presents a compelling arena to reveal the chiral magnetic excitations foreseen to emerge in altermagnets~\cite{Smejkal:2022:2}. 

\begin{figure} 
    \centering
    \includegraphics[width=85mm]{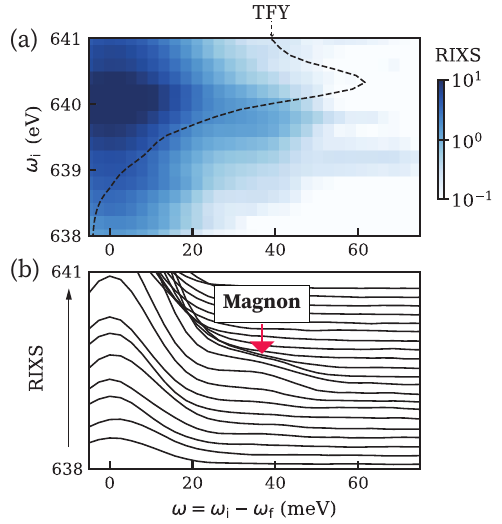}
    \caption{\textbf{Magnetic excitations of MnTe as seen in RIXS.} (a) Incident photon energy $\omega_\mathrm{i}$ dependence of the RIXS signal across the Mn $L_3$ edge. The total fluoresence yield (TFY) is shown as dashed line. (b) Stacked plot highlighting how the magnon shoulder develops slightly off-tuned from the maximum fluorescence signal of the Mn $L_3$ edge, approximately at $\omega \sim 35$ meV. The data were taken at $q = (0.03, 0, 0.5)$ r.l.u.}
    \label{fig:Fig1_2}
\end{figure}

In real space, these magnons can be visualized as spin rotations which precess out-of-phase with neighbors in adjacent sublattice planes, preserving net-zero magnetization [\figureref{fig:Fig1}{(c)}, see Movie~S1 for details]. Due to the distinct symmetry of the two spin sublattices, the spin polarization acquires a spatial texture. In a momentum-space perspective, as schematically shown in \figureref{fig:Fig1}{(d)}, this spatial texture manifests as a spin polarization that varies across the Brillouin zone, consistent with $g-$wave symmetry. These symmetry constraints enforce nodal lines and promote chiral magnetic excitations in well-defined sectors of the Brillouin zone.
\section{Methods}

\subsubsection*{Materials}
Single crystals of MnTe were grown by the Tin-flux method. Mn pieces (Thermo scientific; 99.9\%), Te shots (Thermo scientific; 99.999\%), and Sn shots (Thermo scientific; 99.9999\%) were loaded in a 5-ml aluminum oxide crucible in a molar ratio of 1:1:20. The crucible was sealed in a fused silica ampule under vacuum and heated to 960$^\circ$C over 10 h, homogenized at 960$^\circ$C for 12 h, and then cooled to 840$^\circ$C over 100 h. After reaching 840$^\circ$C the excess flux was decanted from the crystals using a centrifuge leaving behind well-faceted shinny multiple hexagonal single crystals with few millimeters in dimensions.

\subsubsection*{XAS/RIXS}

X-Ray Absorption Spectroscopy (XAS) and Resonant Inelastic X-ray Scattering (RIXS) measurements were performed at the ID32 beamline of the European Synchrotron Research Facility (ESRF) using left- and right-handed circularly polarized light\cite{Brookes:2018}. The combined energy resolution (beamline and spectrometer) was $\sim 21\,\mathrm{meV}$ at the Mn L$_3$ edge (640~eV). As described in the text, the incident photon energy was tuned to optimize the feature in the inelastic spectrum at $\omega \sim 35\,\mathrm{meV}$. Both the sample (using the four-circle high-precision goniometer) and the scattering arm were moved to measure different points in momentum space. The momentum dependent measurements were conducted at (0.1, 0.1, 0.5) r.l.u. and along the (0.2, 0, $\ell$) r.l.u. and ($h$, 0, 0.5) r.l.u. directions with units given in reciprocal lattice units (r.l.u.) of the hexagonal unit cell, having lattice parameters $a = b = 4.17$~\AA ~and $c = 6.75$~\AA. The sample temperature was set to 30\,K.

\begin{figure}
    \centering
    \includegraphics[width = 85mm]{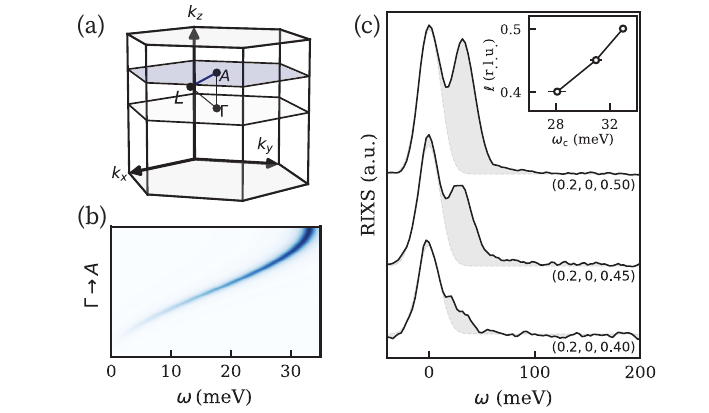}
    \caption{\textbf{Magnetic dispersion along $\mathbf{k_z}$.} (a) Hexagonal Brillouin zone of MnTe, with the triangle spanned by the points $\Gamma$, $A$ and $L$ indicating the region of interest in this work. (b) Illustration of the magnon dispersion of MnTe for the $\Gamma \rightarrow A$ direction from a generic Heisenberg model. (c) RIXS spectra at indicated momentum points showing the shoulder at (0.2, 0, 0.4) r.l.u. developing into a sharp peak for larger momentum (indicated by the grey area). The in-plane momentum was off-tuned along $h$ from specular reflection to reduce the elastic line intensity [see figure~\ref{fig:suppl_fig1}]. The inset shows the peak positions $\omega_\mathrm{c}$ of the magnetic excitation derived from a fitting routine as shown in figure~\ref{fig:suppl_fig2c_fits}. }
    \label{fig:Fig2}
\end{figure} 

\section{Results and Discussion}

To capture chiral excitations experimentally, circular dichroism resonant inelastic X-ray scattering (CD-RIXS) has developed into a promising tool~\cite{Miyawaki:2017, Schueler:2023, Ueda:2023, Ghosh:2023}, more recently also for the investigation of spin-waves~\cite{Zhang:2024}. In our study, circularly polarized photons were used to excite electrons from the Mn $2p$ core levels to its half-filled $3d$ orbitals [\figureref{fig:Fig1}{(e)}] whose weak crystal field splitting is confirmed by both DFT calculations and simulated X-ray absorption spectra [\autoref{fig:suppl_fig_pDOS} and \autoref{fig:simsXAS}]. Within the lifetime of the core-excited intermediate state, energy transferred from the incident photons can create magnetic excitations which manifest as red-shifted, i.e. inelastic, features upon spectrally resolving the scattered photon energy~\cite{Ament:2011}. To locate a clean resonance condition of the magnetic excitation, we tuned the incident photon energy $\omega_\mathrm{i}$ across the Mn $L_3$ absorption edge. Indeed, such an incident photon energy dependence revealed a feature in the inelastic spectrum appearing just below the fluorescence main peak energy [\figureref{fig:Fig1_2}{(a)}]. This excitation is located at an energy transfer of $\omega \sim 35\,\mathrm{meV}$ [\figureref{fig:Fig1_2}{(b)}], an energy scale that aligns with the previously reported magnon band in MnTe at this momentum transfer~\cite{Szuszkiewicz:2005,Liu:2024}. 

To further corroborate the magnetic origin, we measured its momentum dependence along a trajectory in the Brillouin zone [\figureref{fig:Fig2}{(a)}]. In hexagonal anti-ferromagnets with anti-parallel spin ordering along the $c-$axis, the magnon energy typically increases along the $(0,0,\ell)$ direction, exhibiting a local maximum at (0,0,0.5) reciprocal lattice units (r.l.u.) at the $A-$point [\figureref{fig:Fig2}{(b)}], as is the case for MnTe in which it reaches a value of $\sim 32\,\mathrm{meV}$~\cite{Hennion:2002,Szuszkiewicz:2006, Liu:2024}. Conversely, the magnon mode disperses to zero as it approaches the $\Gamma$ point [\figureref{fig:Fig2}{(b)}], reflecting the broken spin-rotational symmetry~\cite{Goldstone:1962}. This behavior also holds for MnTe where the magnon shifts away from the elastic line as the out-of-plane momentum component is tuned towards the $A-L$ plane [see \figureref{fig:Fig2}{(c)} and \autoref{fig:suppl_fig1}]. The magnetic origin of this excitation is further supported by its pronounced temperature dependence [ \autoref{fig:suppl_CAM_temp}] which shows the inelastic signal steadily weakening with increasing temperature, eventually vanishing above the Néel temperature ($T_\mathrm{N} \sim 310\,\mathrm{K}$). This behavior confirms its connection to long-range anti-ferromagnetic order.

\begin{figure*}
    \centering
    \includegraphics[width = \textwidth]{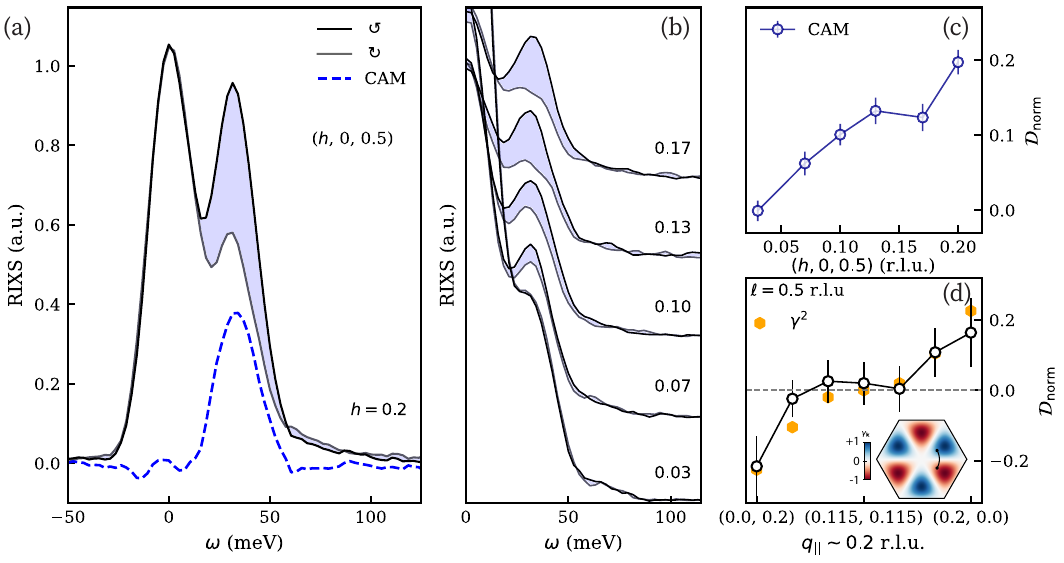}
    \caption{\textbf{Evidence of chiral altermagnons.} (a) Spectra of the magnetic excitations for the two orthogonal circular polarization channels at ($h$, 0, 0.5) rlu with $h=0.2$. The subtraction of the signals in the two channels demonstrates the chiral nature of magnons in MnTe. (b) Momentum dependence ($h$, 0, 0.5) of the chiral altermagnon (CAM) which vanishes towards the high-symmetry $A-$point at (0, 0, 0.5). At high-symmetry points, the modes with opposite chirality are degenerate leading to equal RIXS intensities using circularly polarized light. (c) Intensity dispersion of the CAM where the normalized dichroism $\mathcal{D}_\mathrm{norm} = \left(\mathcal{A}_\circlearrowleft- \mathcal{A}_\circlearrowright\right)/\left(\mathcal{A}_\circlearrowleft + \mathcal{A}_\circlearrowright\right)$ increases systematically towards the center of the $A-L$ direction. (d) $\mathcal{D}_\mathrm{norm}$ for a constant momentum arc in the plane perpendicular to $(0,0,\ell)$ showing the sign change of $\mathcal{D}_\mathrm{norm}$ along with the expected nodal region when approaching the $A-H$ direction. The color map in the lower right corner shows $\gamma_\mathbf{k}$ in $A_{2g}$ symmetry. See figure \ref{fig:vertex} for more details.}
    \label{fig:Fig3}
\end{figure*}

With the magnetic character of the excitation established, we turn to its chiral properties. The spin polarization in momentum space representation [\figureref{fig:Fig1}{(d)}] indicates a significant anisotropy in chiral magnon intensity, shaped by the underlying altermagnetic symmetry. This structure naturally suggests a directional dependence of the dichroic signal: circular dichroism should emerge in regions where the spin polarization is finite, and vanish along nodal directions where the symmetry constraints force it to be zero. 

Towards the $A-L$ direction at (0.2, 0, 0.5) r.l.u., a pronounced difference of the magnon intensity is seen for the two polarization channels, as depicted in \figureref{fig:Fig3}{(a)}. This imbalance reflects the orbital angular momentum transferred during the excitation process, providing a direct measure of the magnon chirality. In this context, the momentum-dependent spin-polarization characteristic for altermagnets~\cite{Smejkal:2022:2} leads to an effective sub-lattice decoupling which manifests as selective sensitivity to one helicity of the circularly polarized X-rays, a hallmark of their fundamentally broken time-reversal symmetry. 

Subtracting these two spectra yields a peak-like signature that encapsulates the chiral portion of the magnetic excitations. We term this excitation a chiral altermagnon (CAM). As the momentum transfer is varied from (0.2, 0, 0.5) r.l.u. towards the $A-$point at (0, 0, 0.5) r.l.u., the CAM intensity drops [\figureref{fig:Fig3}{(b)}], with the normalized dichroism $\mathcal{D}_\mathrm{norm}$ systematically decreasing with smaller in-plane momentum transfer [\figureref{fig:Fig3}{(c)}]. Mapping the CAM along a constant-momentum arc across a Brillouin zone sheet at $\ell=0.5\,$r.l.u. reveals a striking symmetry-driven structure, as $\mathcal{D}_\mathrm{norm}$ undergoes a sign change across a pronounced nodal region in which the dichroic signal vanishes [\figureref{fig:Fig3}{(d)} and \autoref{fig:supplement_dichroism_sample2}]. 

The connection between the observed dichroism and the symmetry of the magnetic excitation arises from how the RIXS cross section transforms under the point group symmetry of the crystal. In MnTe, circularly polarized light couples naturally to the components of the scattering tensor that transforms as the $A_{2g}$ irreducible representation of the $D_{6h}$ point group. This symmetry channel is associated with a distinct set of crystal harmonics whose momentum dependence governs the spatial distribution of chiral intensity in the Brillouin zone [see Supplementary Materials for details]. 

We note that along the $A - L$ direction, no resolvable energy splitting was observed in prior inelastic neutron scattering measurements~\cite{Liu:2024}. However, the presented CD-RIXS data show distinct line-widths for the two polarization channels [Figure~\ref{fig:width}], implying different lower bounds for the magnon lifetimes. This observation may point to a small energy splitting of the magnon branches. Alternatively, the contrast in line-widths could reflect chirality-dependent damping -- a more exotic scenario that would imply a symmetry-allowed but unconventional broadening mechanism. While both scenarios are intriguing, a subtle lifting of the degeneracy currently provides the most straightforward interpretation. Future studies using high-resolution RIXS or polarized neutron scattering will be essential to confirm this picture.

We have demonstrated that chiral altermagnons arise in MnTe, an anti-ferromagnet that displays characteristics canonically associated with finite magnetization. The pronounced dichroic signal whose momentum-dependence aligns with the altermagnetic $g-$wave symmetry underscores how broken spatial symmetries may foster chirality in otherwise collinear spin systems. This holds also if spin-orbit coupling is taken into account~\cite{Belashchenko:2025}. Moreover, our results complete a crucial triad in the rapidly developing field of altermagnetism: the observations of spin-split electronic band structures~\cite{Krempasky:2024, Lee:2024, Osumi:2024, Hajlaoui:2024}, the splitting of magnon bands~\cite{Liu:2024}, and now, the direct experimental observation of chirality in magnetic excitations, which positions MnTe as the archetypal altermagnet. 

Future investigations will need to address the role of chiral magnons in shaping both transport and spectroscopic properties of altermagnetic materials. Advancing this field further will require a robust theoretical framework that captures the momentum-resolved nature of these excitations and their interaction with electronic and magnetic degrees of freedom, including damping-driven effects. 

Lastly, as altermagnets lack stray magnetic fields, they are ideal for compact and interference-free device integration. By demonstrating that altermagnets host chiral excitations, our work paves the way for novel spin-based applications in which chiral altermagnons take center stage.


\newpage

\section*{Acknowledgments}
\paragraph*{Funding:}
This work was supported by the Department of Energy, Basic Energy Science. The experiment was performed at the European Synchrotron Radiation Facility in Grenoble, France under proposal HC-5866, DOI: 10.15151/ESRF-ES-1901557821. MS acknowledges support by the Munich Quantum Valley, which is supported by the Bavarian state government with funds from the Hightech Agenda Bayern Plus.
\paragraph*{Author contributions:}
D.J. conceived the study, led the experiment and analyzed the data. D.J., S.S.S., M.S., M.N., R.N. and S.G. performed the experiments. R.B.R. and N.J.G. synthesized and characterized the samples. E.G.L. performed the DFT and multiplet calculations. D.J., R.B.R, E.G.L., S.S.S., M.S., M.N., R.N., F.Y., K.K., N.B., L.S., G.L.D., N.J.G., S.G., and M.F.K. discussed and interpreted the results and contributed to the final version of the manuscript.
\paragraph*{Competing interests:}
There are no competing interests to declare.
\paragraph*{Data and materials availability:}
Data underlying the paper are permanently deposited at \url{10.15151/ESRF-ES-1901557821} with a current embargo until 2027. We assure public availability, once the paper is published. 

\bibliography{MnTe_bib}

\clearpage
\newpage 
\onecolumngrid

\setcounter{figure}{0}
\clearpage
\newpage
\subsection*{Supplement}

\renewcommand{\thefigure}{S\arabic{figure}}
\renewcommand{\thetable}{S\arabic{table}}
\renewcommand{\theequation}{S\arabic{equation}}
\renewcommand{\thepage}{S\arabic{page}}
\setcounter{figure}{0}
\setcounter{table}{0}
\setcounter{equation}{0}
\setcounter{page}{1} 


\subsubsection*{DFT}
Spin-polarized Density Functional Theory (DFT) calculations of the projected Density of States (pDOS) of MnTe and MnO were done using the Quantum ESPRESSO (v7.4) code ~\cite{Giannozzi_QEspresso}. Pseudopotentials were generated using the code Optimized Norm-Conservinng Vanderbilt PSeudopotential (ONCVPSP)
 scalar-relativistic version 3.3.1~\cite{Hamann_ONCVPSP}. The generalized gradient approximation (GGA) of Perdew-Burke-Ernzerhof (PBE) with a rotationally invariant Hubbard U and J as implemented by Liechtenstein et al. was employed~\cite{Perdew_PBE,Liechtenstein_LDAU}. For the U and J values, we used 4.0 and 0.97 eV, respectively, as done in previous MnTe DFT calculations~\cite{Kriegner:2017}. We used the same values for MnO, for consistency. Layered antiferromagnetic ordering was used in both materials. Tetrahedral smearing with a planewave cut off of 130 Ry and a uniform $k$-grid of 16x16x8 for MnTe and 12x12x12 for MnO were used. The projection of the Kohn-Sham states was done on the orthogonalized atomic orbitals that are rotated to the basis in which the atomic occupation matrix is diagonal.

\subsubsection*{Multiplet}
Multiplet calculations followed the formalism of our previous work~\cite{Jost:2025}. A single site of one metal atom ($3d$ orbitals), and 3 ligands ($2p$ orbitals each) was employed. The single site is centered at (\begin{math}\pi/2,\pi/2\end{math}) in momentum. The Hamiltonian for the multiplet cluster can be expressed as:
\begin{align}
    \hat H &= 
     \frac{1}{2}\sum_{i,\sigma,\sigma'} \sum_{\mu,\nu,\mu',\nu'} U_{\mu,\nu,\mu',\nu'} \hat c^\dagger_{i,\mu,\sigma}\hat c^\dagger_{i,\nu,\sigma'}\hat c_{i,\mu',\sigma'}\hat c_{i,\nu',\sigma} \nonumber\\ 
     & + \sum_{i,j,\sigma}\sum_{\mu,\nu} t_{i,j}^{\mu,\nu} \hat c^\dagger_{i,\mu,\sigma}\hat c_{j,\nu,\sigma} \nonumber + \sum_{i,\mu,\nu,\sigma}V_{CEF}(\mu,\nu)c^\dagger_{i,\mu,\sigma}\hat c_{i,\nu,\sigma}\nonumber \\
& + \frac{1}{2}\sum_{i,\sigma,\sigma'} \sum_{\mu,\nu,\mu',\nu'} U_{\mu,\nu,\mu',\nu'} \hat c^\dagger_{i,\mu,\sigma}\hat d^\dagger_{i,\nu,\sigma'}\hat c_{i,\mu',\sigma'}\hat d_{i,\nu',\sigma} \nonumber\\
& - \sum_{i,\sigma,\sigma'} \sum_{\mu,\nu} \lambda_{\mu,\nu}^{\sigma,\sigma'} \hat d^\dagger_{i,\mu,\sigma} \hat d_{i,\nu,\sigma'} + \sum_{i}\Delta_{i}n_{i}
\end{align}

Where $i$, $j$ refer to the different atomic sites, $\mu$, $\nu$ refer to different sets of $l$, $m$ quantum numbers, and $\sigma$ refers to spin. The creation and annihilation operators denoted in $c$ are valence electron (hole) operators; whereas operators denoted in $d$ are core electron (hole) operators. 

The first term includes a Hubbard-like $U$ term for the coulomb direct and exchange interactions for transition-metal oxides. The second term includes a \textit{t} hopping element between different atomic sites and their orbitals. The third term includes an octahedral crystal field ($V_{CEF}$) for the $d$-orbitals in the metal atom, the fourth term is the core-valence coulomb interaction, the fifth term is the spin-orbit coupling $\lambda$ at the core and the last term is the charge transfer energy $\Delta$ at each atomic site. The multi-particle eigenstates for a Hamiltonian of an \textit{N} hole cluster and one for a \textit{N-1} hole cluster with a core hole serve as the initial (\textit{i}), intermediate ($\nu$), and final (\textit{f}) states for the calculation of X-ray absorption spectra by Fermi's golden rule:
\begin{align}
&\kappa_{e_i, k_i}(\omega)=
\frac{1}{\pi Z} \sum_{i,\nu} e^{-\beta E_i} \mid \langle\nu\mid \hat D_{k_i}({e_i})\mid i \rangle\mid^2 \delta(\omega-(E_\nu-E_i))
\end{align}
The parameters of our calculations can be found in Table \ref{tab:mult_params}. To account for finite lifetime effects, a Lorentzian broadening of 0.1 eV at full width half maximum was applied to the spectral function.

\subsubsection*{A\textsubscript{2g} cross section}

The relationship between chirality and the RIXS matrix elements can be understood from group-theoretical considerations. In general, the RIXS intensity is governed by the Kramers-Heisenberg expression \cite{Ament:2011}:

\begin{equation}
    \mathcal{R}(\omega_\mathbf{k},\omega_{\mathbf{k}^\prime}, \mathbf{k}, \mathbf{k}^\prime, \mathbf{e}, \mathbf{e}^\prime) \propto \sum_f \left| \sum_\nu \frac{\langle f | \mathcal{D}^{\prime\dagger} |\nu\rangle \langle \nu | \mathcal{D} | g \rangle}{E_g + \hbar \omega_\mathbf{k} - E_\nu + i \Gamma_\nu} \right|^2 \delta(E_g - E_f + \hbar\omega),
    \label{eqn:KH}
\end{equation}

where $|g\rangle, |\nu\rangle, |f\rangle $ are the ground, intermediate, and final states with energies $ E_g, E_\nu, E_f $, respectively. The transition operator $ \mathcal{D} = \sum_i \mathbf{e} \cdot \mathbf{p}_i \, e^{i \mathbf{k} \cdot \mathbf{r}_i} $ contains the dependence on the incident polarization $ \mathbf{e} $, momentum $ \mathbf{k} $, and the electronic degrees of freedom. In the dipole limit, this operator simplifies to $ \mathcal{D} = \sum_i \mathbf{e} \cdot \mathbf{r}_i \, e^{i \mathbf{k} \cdot \mathbf{R}_i} $, where $ \mathbf{R}_i $ denotes the position of the scattering centers. These matrix elements are thus sensitive to both polarization and momentum transfer. 

The symmetry properties of the scattering cross section are constrained by the transformation of the dipole operators. In the $D_{6h}$ point group, the dipole operator transforms as $E_{1u}$ for in-plane polarization and $A_{2u}$ for out-of-plane polarization. The tensor product of two operators thus yields: 
\begin{equation}
E_{1u} \otimes E_{1u} = A_{1g} \oplus A_{2g} \oplus E_{2g}, \qquad
A_{2u} \otimes A_{2u} = A_{1g}, \qquad
E_{1u} \otimes A_{2u} = E_{2g}.
\end{equation}
This decomposition shows that all three even-parity irreducible representations $A_{1g}, A_{2g}, $ and $E_{2g}$ are in principle accessible, depending on the incident and outgoing polarization configuration. 

If the outgoing polarization is not analyzed as is the case for most RIXS experiments, then all symmetry channels compatible with the incoming polarization are retained. For instance, with in-plane linearly polarized light,  $A_{1g}, A_{2g},$ and $E_{2g}$ channels contribute to the cross section. Notably, when employing circular polarization states, the tensor product allows for mixtures of the isotropic $A_{1g}$ and antisymmetric $A_{2g}$ term, the latter of which couples to the $\sigma_2$ Pauli matrix. In other words, these states are naturally accessed when using circularly polarized X-rays, which inject finite angular momentum and selectively probe the antisymmetric components of the scattering tensor. 

A direct connection to our experiment can be drawn, when considering the momentum dependence of the scattering vertex. Each symmetry channel is associated with a distinct set of crystal harmonics that define how the corresponding excitation intensity varies across the Brillouin zone. These crystal harmonics serve as fingerprints of the symmetry of the underlying excitation ~\cite{Hayes:1979}, in our case a magnon. For $A_{2g}$ symmetry, the vertex $\gamma_\mathbf{k}$ changes sign according to Figure \figureref{fig:vertex}{(a)} across different portions of the Brillouin zone, thereby emphasizing the corresponding chiral branch of the magnetic excitation. The measured RIXS intensity scales as $\gamma_\mathbf{k}^2$, shown in Figure \figureref{fig:vertex}{(b)}. 

This vertex structure naturally explains the momentum-resolved pattern observed in the CD-RIXS measurements. The normalized dichroism $\mathcal{D}_\mathrm{norm} = (\mathcal{A}_{\circlearrowleft} - \mathcal{A}_\circlearrowright)/(\mathcal{A}_{\circlearrowleft} + \mathcal{A}_\circlearrowright)$ tracks the intensity modulation of $\gamma_\mathbf{k}^2$ across the Brillouin zone. The observed sign change in $\mathcal{D}_\mathrm{norm}$ as the momentum transfer sweeps across symmetry-equivalent directions provides direct evidence of an underlying $A_{2g}$ symmetry, strongly supporting the identification of the excitation as a chiral altermagnon.

\newpage


\begin{figure} 
	\centering
	\includegraphics{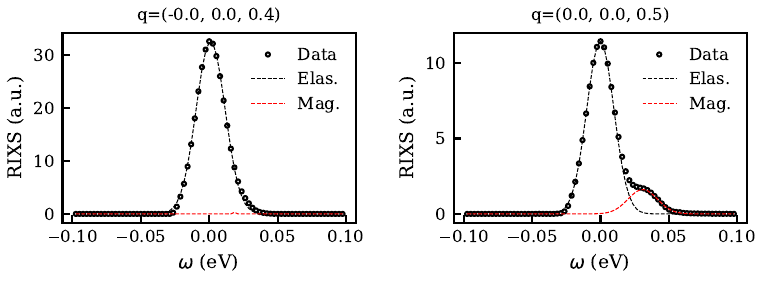} 
	\caption{\textbf{Momentum dependence along the $\Gamma - A$-direction.}
    RIXS measurement at (0, 0, 0.4)~r.l.u. and (0, 0, 0.5)~r.l.u. showing the magnetic excitation off the elastic line at the $A-$point. The strong elastic contribution is due to specular reflection. }
	\label{fig:suppl_fig1} 
\end{figure}

\begin{figure} 
	\centering
	\includegraphics{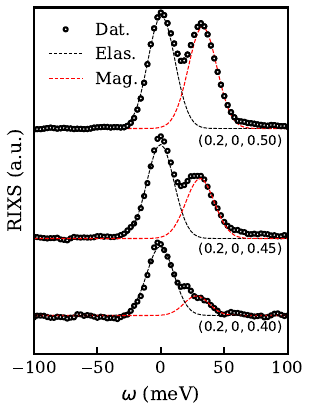} 
	\caption{\textbf{Fitting results of the data presented in figure \ref{fig:Fig2}.} The data was fitted using two Gaussian components. The black dashed line represents the fit to the elastic line, whereas the red dashed line takes into account the magnon line-shape. The error bars in figure \ref{fig:Fig2} were derived from the covariance matrix of the fit.}
	\label{fig:suppl_fig2c_fits}
\end{figure}

\begin{figure} 
	\centering
	\includegraphics[width=85mm]{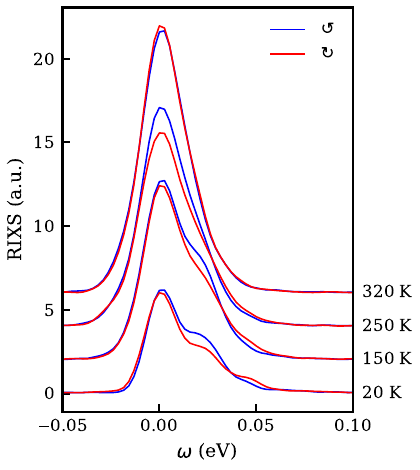} 
	\caption{\textbf{Temperature dependence of the magnetic excitation for both left- and right- circularly polarized X-rays.} The magnetic excitation is clearly seen at 20\,K and develops into a shoulder upon increasing the temperature to 150\,K. At 250\,K, the elastic line exhibits an asymmetry, presumably from the magnon energy softening. }
	\label{fig:suppl_CAM_temp} 
\end{figure}

\begin{figure} 
	\centering
	\includegraphics[width=85mm]{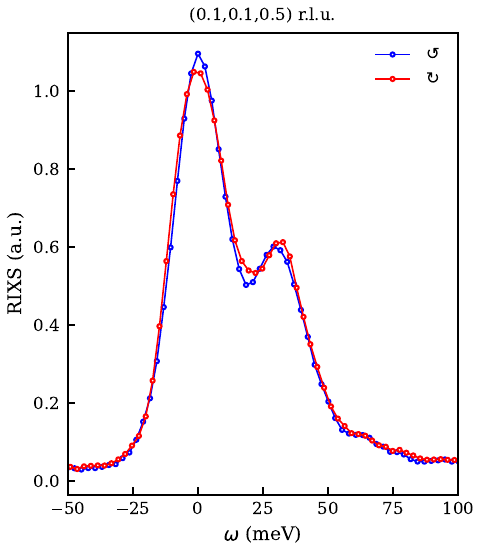} 

	\caption{\textbf{Circular dichroism measurement along the nodal direction.}
    CD-RIXS measurement at (0.1, 0.1, 0.5) r.l.u. showing a strongly suppressed dichroism compared to an equivalent momentum point along the $A-L$-direction.}
	\label{fig:suppl_fig2} 
\end{figure}

\begin{figure}[h!]
    \centering
    \includegraphics[width= 1.0\textwidth]{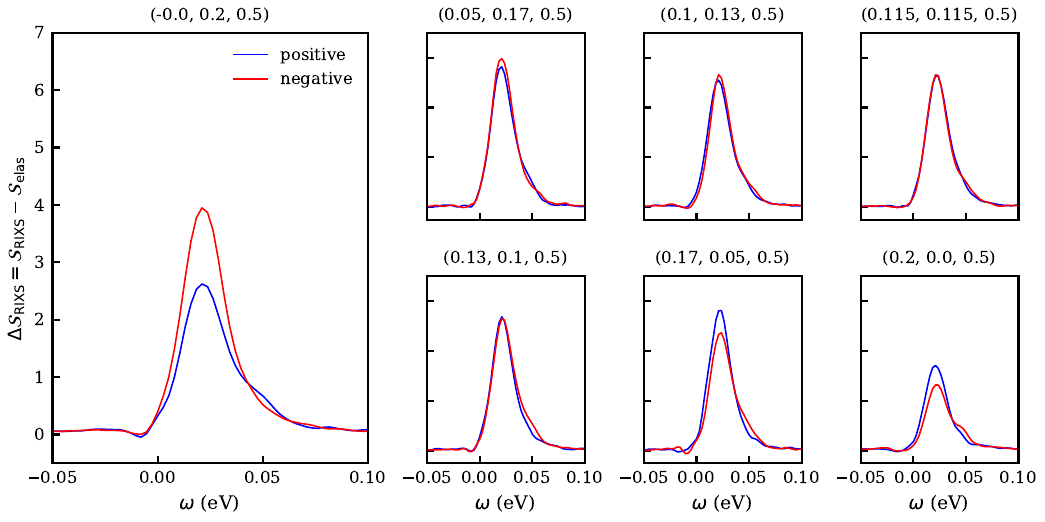}
    \caption{CD-RIXS across a constant momentum arc from (0.2, 0, 0.5) to (0, 0.2, 0.5) r.l.u.}
    \label{fig:supplement_dichroism_sample2}
\end{figure}

\begin{figure}
    \centering
    \includegraphics{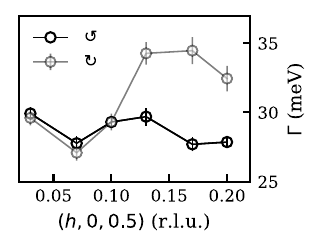}
    \caption{Line width of the chiral altermagnon (CAM) from figure \ref{fig:Fig3}.}
    \label{fig:width}
\end{figure}

\begin{figure}
    \centering
    \includegraphics[width = 1.0\textwidth]{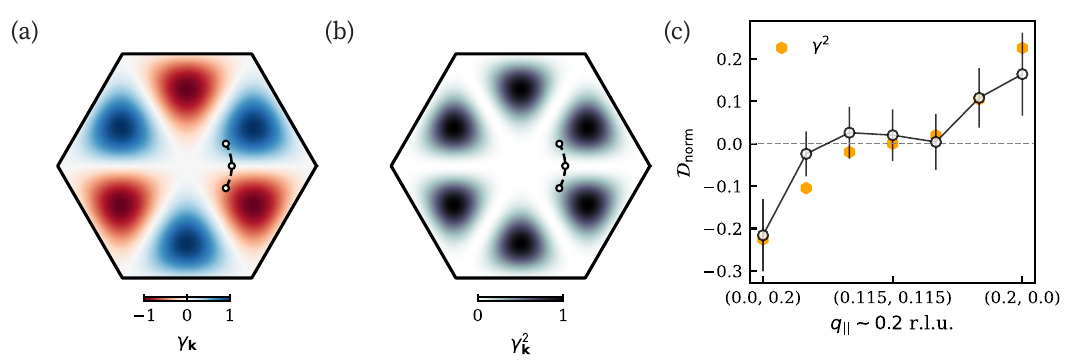}
    \caption{\textbf{Comparison of the scattering amplitude and the momentum dependent normalized dichroism $\mathcal{D}_\mathrm{norm}$.} (a) Momentum dependent light scattering vertex $\gamma_\mathbf{k}$ for $A_{2g}-$ symmetry. The dashed line shows the constant momentum arc along which the RIXS data of figure \ref{fig:supplement_dichroism_sample2} was taken. (b) Squared light scattering vertex $\gamma_\mathbf{k}^2$ reflecting the fundamental two-photon RIXS process. (c) Comparison of the normalized dichroism of the chiral altermagnon with the squared scattering vertex, showing $\mathcal{D}_\mathrm{norm} \sim \gamma_\mathbf{k}^2$. Note:  Negative values of $\mathcal{D}_\mathrm{norm}$ result from convention of the subtraction procedure. The sign of $\gamma_\mathbf{k}^2$ is flipped in the corresponding momentum range to accommodate that convention.}
    \label{fig:vertex}
\end{figure}

\begin{figure} 
	\centering
	\includegraphics[width=160mm]{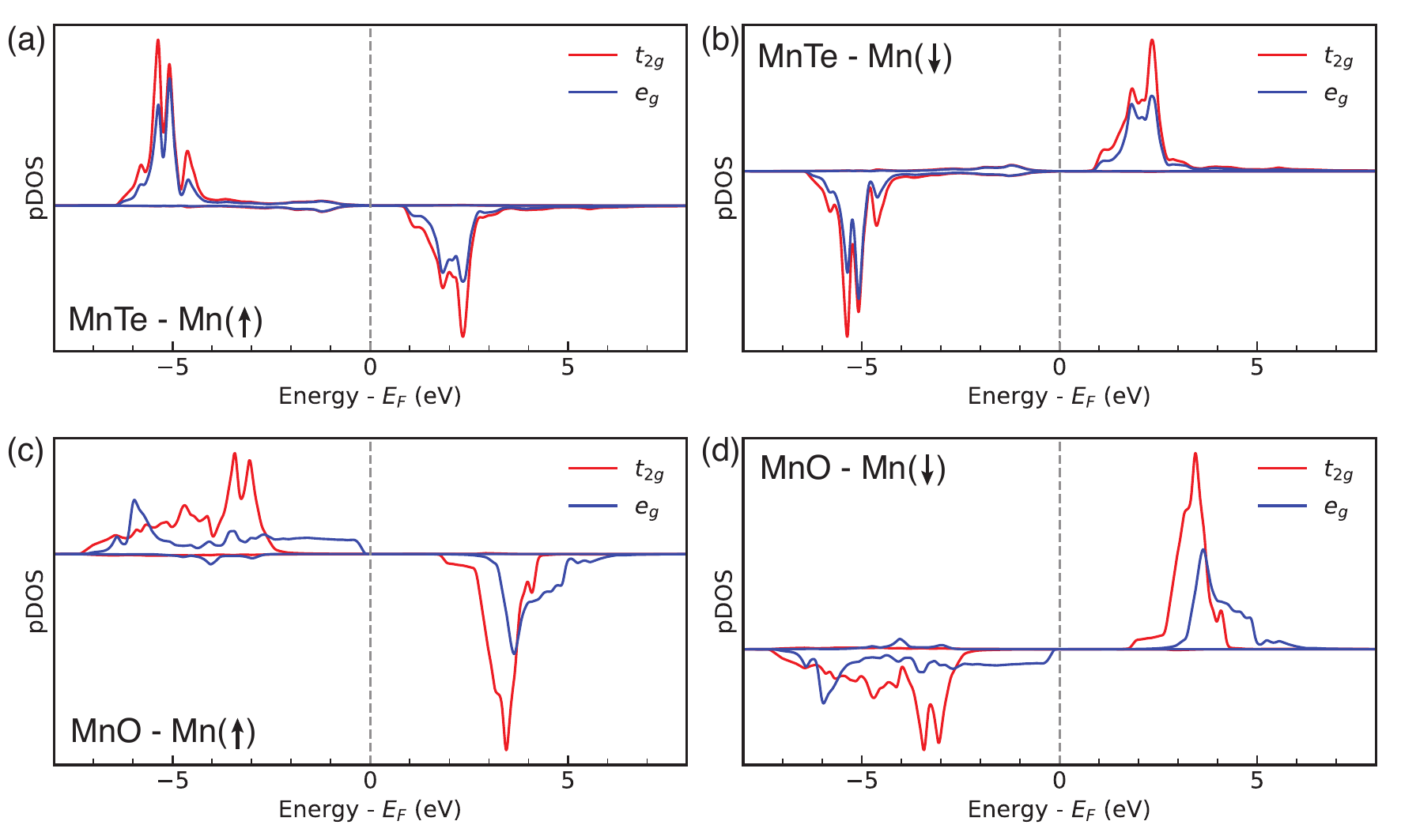}

	\caption{\textbf{pDOS comparison of MnTe and MnO Mn $d$ states.} DFT calculations show the minimal energy splitting of $t_{2g}$ and $e_g$ manifolds in MnTe, compared to MnO, which is known to have $d$-orbital degeneracy due to octahedral coordination. }
	\label{fig:suppl_fig_pDOS} 
\end{figure}

\begin{figure}
    \centering
    \includegraphics{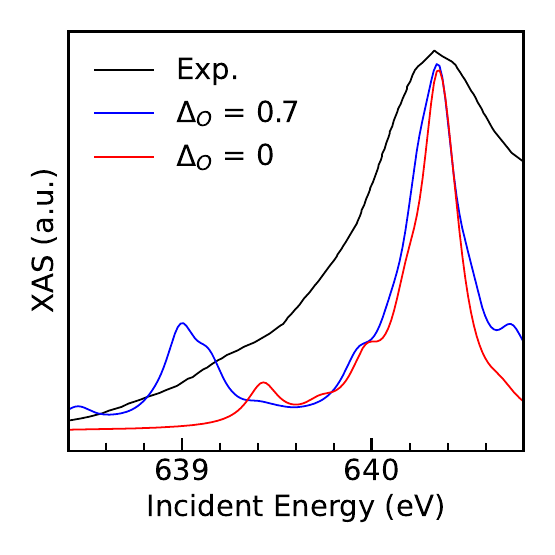}
    \caption{\textbf{Multiplet calculation for the X-ray absorption spectra using different crystal field energies.} The experimental results (black) lack the pre-peak that naturally arises for a weak crystal field splitting ($\Delta_O = 0.7$). In contrast, the XAS multiplet structure is well captured for a vanishing $\Delta_O = 0$.}
    \label{fig:simsXAS}
\end{figure}

\begin{table}
	\centering
	\caption{\textbf{Multiplet calculation parameters and orbital occupations for MnTe and MnO effective models.} Slater-Condon parameters ($F$ and $G$) and spin-orbit interactions (${\zeta}$) were set by Hartree-Fock values of Mn$^{2+} (d^5)$~\cite{Haverkort:2005_thesis}. $U_{dd}$ and (${\Delta}$) were set to recover a positive charge transfer ground state (majority $d^5$). The crystal field ($\Delta_O$) and hybridization values ($t_{t_{2g}}/t_{pp}$) for MnO were set to $x_1=0.7$ and $x_2/x_3=0.6/0.1$, respectively, setting a strong crystal and ligand fields which significantly split the $d$-orbital degeneracy. For MnTe, these were set to $x_1=0$ and $x_2/x_3=0.2/0$, respectively, having no crystal field and a weak ligand field, resulting in near degenerate $d$-orbitals, as seen in our DFT results. All units are in eV.\\}
    \renewcommand{\arraystretch}{1.1}
    \setlength{\tabcolsep}{2pt}
	\label{tab:mult_params}
	\begin{tabular}{|c|ccccccccccccc|}
    \hline
    \multicolumn{14}{|c|}{\textbf{CTFHAM parameters for MnL\textsubscript{3} cluster}} \\\hline
    Config &$U_{dd}$ & $\Delta_O$ & $\Delta$ & $\zeta_{3d}$ & $F_{dd}^{2}$ & $F_{dd}^{4}$ & $\zeta_{2p}$ & $F_{pd}^{2}$ & $G_{pd}^{1}$ & $G_{pd}^{3}$ & $t_{t_{2g}}$ & $t_{e_g}$ & $t_{pp}$ \\\hline
     $2p^63d^5$ & 4.5 & $x_1$ & 4.5 & 0.040 & 10.315 & 6.413 & - & - & - & - & $x_2$ &  0.4$x_2$ & $x_3$ \\\hline
     $2p^53d^6$ & " & " & " & 0.053 & 11.154 & 6.942 & 6.846 & 6.320 & 4.603 & 2.617 & " & " & " \\\hline
     \multicolumn{14}{|c|}{\textbf{Ground state hole orbital occupations}} \\\hline
     \multicolumn{2}{|c|}{Material} & $d_{x^2-y^2}$ & $d_{z^2}$ & $d_{xy}$ & $d_{xz}$ & $d_{yz}$ & \multicolumn{7}{|c|}{Contributions to $|\Psi_{GS}|^2$} \\\hline
     \multicolumn{2}{|c|}{$\mathrm{MnO}$} & 0.92 & 0.93 & 0.98 & 0.98 & 0.98 & \multicolumn{7}{|c|}{$0.81 \langle d^5\rangle + 0.18 \langle d^6L\rangle + 0.01 \langle d^7L^2\rangle$} \\\hline
     \multicolumn{2}{|c|}{$\mathrm{MnTe}$} & 0.99 & 0.99 & 1.0 & 1.0 & 1.0 & \multicolumn{7}{|c|}{$0.97 \langle d^5\rangle + 0.03 \langle d^6L\rangle$} \\\hline
	\end{tabular}
\end{table}

\end{document}